\documentclass[journal,comsoc]{IEEEtran}

\usepackage[T1]{fontenc}
\def\baselinestretch{1}

\ifCLASSINFOpdf
\else
\fi

\usepackage{amsmath}
\usepackage[cmintegrals]{newtxmath}
\usepackage{algorithm}
\usepackage{algorithmic}
\usepackage{mathtools}
\usepackage{subcaption}

\usepackage{stackengine}

\newcommand{\ab}{{\textbf{a}}}

\usepackage{lipsum}
\usepackage{fancyhdr}
\usepackage{array}
\usepackage{xcolor}

\begin{document}

	\title{
		Distributed Estimation with Partially Accessible Information: An IMAT Approach to LMS Diffusion}
	
			\author{Mahdi Shamsi and Farokh Marvasti,~\IEEEmembership{Life~Senior~Member,~IEEE}%
}
	\maketitle
	
		\begin{abstract}
			Distributed algorithms, particularly Diffusion Least Mean Square, are widely favored for their reliability, robustness, and fast convergence in various industries. However, limited observability of the target can compromise the integrity of the algorithm. To address this issue, this paper proposes a framework for analyzing combination strategies by drawing inspiration from signal flow analysis. A thresholding-based algorithm is also presented to identify and utilize the support vector in scenarios with missing information about the target vector's support. The proposed approach is demonstrated in two combination scenarios, showcasing the effectiveness of the algorithm in situations characterized by sparse observations in the time and transform domains.
	\end{abstract}
	
	\IEEEpeerreviewmaketitle
		
	\section{Introduction}
	\label{sec:format}
Distributed algorithms have gained substantial attention across various industries, including weather prediction \cite{nassif2020learning}, federated learning \cite{kairouz2021advances}, and multi-agent reinforcement learning \cite{zhang2021multi}. This preference can be attributed to their robustness, reliability, and faster convergence rates in comparison to low-cost processing units \cite{nassif2020learning}. One such algorithm, Diffusion Least Mean Square (DLMS) \cite{cattivelli2009diffusion,abdolee2014diffusion,chen2015diffusion}, utilizes a diffusion process and promotes collaboration amongst multiple nodes to estimate the desired signal by exchanging information and refining individual estimates. Notably, DLMS can handle non-stationary data, making it suitable for diverse applications.

Previous studies have commonly assumed that unrestricted access to the target vector is available to every node, enabling independent estimation of the target vector when sufficient resources are available. Partial diffusion has also been proposed as a means of intentionally reducing communication load, with nodes transmitting subsets of their intermediate estimate vectors \cite{arablouei2013distributed, vahidpour2017analysis}.
The concept of sparse distributed estimation has been examined in \cite{yim2015proportionate}. This framework was leveraged to advance the resilience of systems when confronted with conditions exhibiting sparsity of the target vector.
However, impaired nodes or a non-stationary environment can compromise the integrity of the algorithm due to misleading data and information flowing over the network, it was demonstrated that each node can actively mitigate the effect of malfunctioning nodes on the local estimation by employing an appropriate weighting scheme \cite{shamsi2021flexible} .

Given the previous oversight in related studies, this paper addresses situations where nodes have limited visibility of the target and must cooperatively transmit available information to achieve a consensus on the target vector. To achieve this, the paper proposes a framework for analyzing combination strategies, drawing inspiration from signal flow analysis outlined in \cite{shamsi2021flexible}. Furthermore, this paper investigates scenarios with missing information about the target vector's support and presents a thresholding-based algorithm to identify and utilize the support vector for accurate estimation exchange. Previous studies have shown that an Iterative Method with Adaptive Thresholding (IMAT) can be advantageous in recovering missing samples with unknown sparsity location \cite{shamsi2020nonlinear,marvasti2012unified,zamani2020random,sadrizadeh2020fast,sadrizadeh2020impulsive}.


In a network comprising of N nodes, a single estimator at each node attempts to employ an adaptation procedure at every time step. Following this, it shares its inner estimation, $\vec{\tilde{\omega}}_i(t)$, with its neighbors $\aleph_i$. During the combination phase, each node utilizes the received/shared information to refine its inner estimation and create its own estimation ($\vec{{\omega}}_i(t)$). The present study considers a scenario where each node has only partial access to the target signal and thus, needs to exchange information with its neighbors. 

We provide a mathematical proof in this paper, establishing the existence of a unique optimal solution for such a problem, which necessitates complete knowledge about each node's accessibility to the target signal in terms of their observability vector and noise level of measurements. 

However, it may be impractical to acquire such information in reality. As such, we propose an algorithm that enables each node to efficiently disseminate its inner estimation throughout the network and extract the target support vector, while also allowing control over data flow and combination processes of estimations received from neighbors. 

Our simulations demonstrate that the proposed algorithm successfully mitigates the conventional Diffusion LMS's failure, leading to an improvement in Mean Square Deviation (MSD) by 30-40 dB and achieving performance parity with fully observable target scenarios with an observability ratio.

The structure of the article is as follows. Section \ref{sec:LMSDiff} investigates DLMS with partial observations while proposing an algorithm for disseminating partially observed information across the network. Section \ref{sec:sim} details the proposed approach and offers a demonstration of its performance in two combination scenarios characterized by sparse observations in time and transform domains. The article concludes with Section \ref{sec:con}.
\section{LMS Diffusion with Partial Observations}
\label{sec:LMSDiff}
The primary algorithm employed in this study is the  LMS algorithm, which has demonstrated exceptional performance in previous works \cite{hua2020diffusion}. In an arbitrary node $i$, the local estimation procedure can be summarized as follows:
\begin{equation}
	g(\vec{\tilde{\omega}}_i(t), \Theta_i(t)) = \Bigg\{\begin{array}{cc}
		\text{measurement: }d_i(t) \leftarrow \vec{\ab}_{i,t}^TM_i\vec{\omega}^{opt} + \nu_i(t)\\
		\text{error calculation: }e_i(t) \leftarrow d_i(t) - \vec{\ab}_{i,t}^T\vec{\tilde{\omega}}_i(t-1)\\
		\text{adaptation: }\vec{\omega}_i(t) = \vec{\omega}_i(t-1) + \mu e_i(t) \vec{\ab}_{i,t} \notag
	\end{array}\notag
\end{equation}
where $\Theta_i(t)$ denotes the set of parameters, including the measurement vector $\vec{\ab}_{i,t}$ and the adaptation rate $\mu$. The term $\nu_i(t)$ represents measurement noise.

To execute the combination step, the current methodology operates on the assumption of full access to the target vector ($\vec{\omega}^{opt}$), which is referred to as a fully observable target. However, practical scenarios may not allow such access, limiting nodes to only partial target information obtained through measurement, which is referred to as partially observable targets. In such situations, a mask operator designated as $M_i=T'D_iT$ is employed to depict the sparsity of observability within the domain, where the diagonal components of matrix $D_i$ denote the observability of each component of the target vector in the transform domain\footnote{To prevent any potential ambiguities, we have utilized $T'$ to represent the transpose of matrix $T$, while for matrices such as $A$, we represent the transpose as $A^T$.}.

Appendix \ref{apen:conv} illustrates the inadequacy of the common combination approach in this specific scenario, which calls for an optimal combination approach to solve the problem. In Appendix \ref{apen}, it is emphasized that complete knowledge of the observability vectors is pivotal in deriving the optimal combination weights. Additionally, Appendix \ref{apen:sol} demonstrates that the optimal weights are contingent on the Signal-to-Noise Ratio (SNR) level and the attenuation induced by the measurement.

In practical settings, observability vector information is frequently unavailable, rendering the support and values unknown. It is reasonable to assume that the attenuation is trivial, implying that $D_i(j)\in\{0,1\}$, but having knowledge of the support vector is vital. To address this issue, a thresholding approach inspired by iterative sparse signal recovery techniques is utilized.

As each estimator gradually approaches its local target $M_i\vec{\omega}^{opt}$ after several iterations, we may use the magnitude of its components in the transform domain to estimate its support. By defining $\beta(t) = \beta_1e^{-t/\tau}+\beta_0$ as the thresholding level at time $t$, we consider components with magnitudes greater than $\beta(t)$ as observed values of $\vec{\omega}^{opt}$. This enables us to diffuse information throughout the network.

In this method, each node shares the components of its own estimation in the transform domain that are confirmed by the threshold level. To prevent the propagation of shared information from being hindered, each node is required to share relevant information with its neighbors regarding it received and estimated components, as detailed in Alg. \ref{alg:Dimat}.
\begin{algorithm}
	\begin{algorithmic}
		\FOR{$t=1,\dots,T$}
		\STATE Adaptation, $\vec{{\Omega}}_i = g(\vec{\omega}_i,\Theta_i)$
		\STATE Thresholding, $\tilde{D}_i = \Xi(\vec{{\Omega}}_i,\beta)$
		\STATE Combination, $\vec{\psi}_i, \vec{\omega}_i= h(\vec{\psi}_j,\vec{\tilde{\Omega}}_j,\tilde{D}_j\vert j\in \aleph_i)$
		\ENDFOR 
	\end{algorithmic}
	\caption{IMAT-based LMS Diffusion.}
	\label{alg:Dimat}
\end{algorithm}
The function $\Xi(\vec{x},\beta)$ computes the indicator diagonal matrix by $\tilde{D} \leftarrow \text{diag}(|\vec{x}|>\beta)$.
The function $h(\vec{\psi}_j,\vec{\Omega}_j,{D}_j\vert j\in \aleph_i)$ is determined through a multistep process involving local updating, diffusion, and combination, as follows:
\begin{align}
	\Bigg\{\begin{array}{cc}
		\text{Locally update:}&\hspace{-5mm}\vec{\psi}_i \leftarrow D_i((1-\eta) \vec{\psi}_i + \eta\vec{\Omega}_i)+(\mathbb{I}-D_i)\vec{\psi}_i\notag\\
		\text{Diffuse:}&\hspace{-5mm}\vec{\psi}_i  \leftarrow \sum_{j\in \aleph_i}D_{\text{comb}, j}\vec{\psi}_j\notag\\
		\text{Combine:}&\hspace{-5mm}\vec{\omega}_i  \leftarrow T'[D_i((1-\alpha) \vec{\psi}_i + \alpha\vec{\Omega}_i)+(\mathbb{I}-D_i)\vec{\Omega}_i ]\notag
	\end{array}
\end{align}
Here, $D_{\text{comb}, j}$ is defined by a combination strategy whereby, in instances of equal observability power and noise variance over different nodes, it is considered an average over each active component of the neighboring nodes. 

It should be noted that the parameters $\alpha$ and $\eta$ control the diffusion flow over the network. Specifically, $\alpha$ controls the inflow of information into the node's estimation from the network, while $\eta$ controls the outflow of information from each node through the network. In other words, $\alpha$ allows incoming information from neighboring nodes to refine the estimated components located on the estimated target support, while $\eta$ regulates the extent to which information flows through nodes over the network.

To increase the exchange of information during transient steps, it might be beneficial to set $\eta>0$ and $\alpha=0$. This allows for more fluid communication among nodes. However, during the steady state, it is advisable to restrict the information flow to the combining step by setting $\eta=0$ and using an $\alpha$ value between $0$ and $1$.
\section{Simulation Results and Discussion }
\label{sec:sim}
\begin{figure}[hbt!]
	\centering
	\begin{subfigure}[b]{0.23\textwidth}
		\centering
		\stackunder[5pt]{\includegraphics[width=\textwidth]{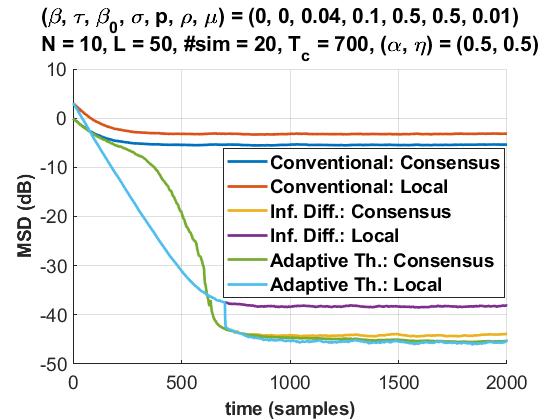}}{\scriptsize{Partialy obseravable.}}
	\end{subfigure}
	\hfill
	\begin{subfigure}[b]{0.23\textwidth}
		\centering
		\stackunder[5pt]{\includegraphics[width=\textwidth]{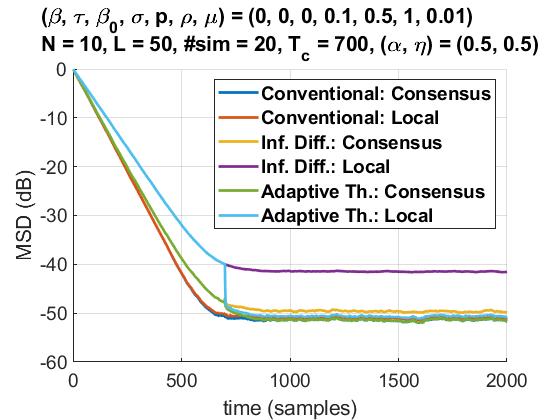}}{\scriptsize{Fully obseravable.}}
	\end{subfigure}
	\caption[font=small]{
		The DLMS algorithm with Time-Domain Observations.
	}
	\label{fig:DE}
\end{figure}

\begin{figure}[hbt!]
	\centering
	\begin{subfigure}[b]{0.23\textwidth}
		\centering
		\stackunder[5pt]{\includegraphics[width=\textwidth]{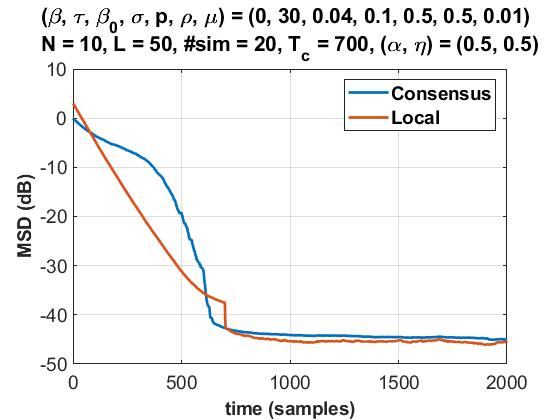}}{\scriptsize{Low noise.}}
	\end{subfigure}
	\hfill
	\begin{subfigure}[b]{0.23\textwidth}
		\centering
		\stackunder[5pt]{\includegraphics[width=\textwidth]{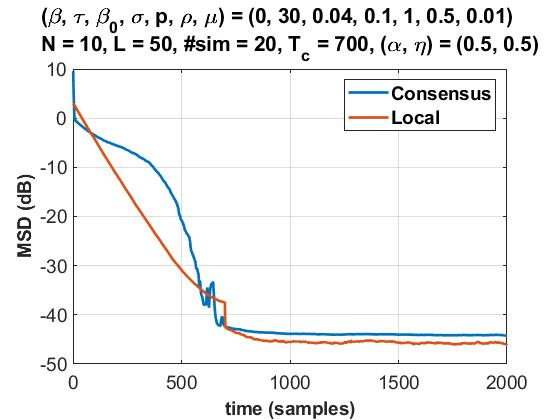}}{\scriptsize{Low noise: Complete graph.}}
	\end{subfigure}
	\hfill
	\begin{subfigure}[b]{0.23\textwidth}
		\centering
		\stackunder[5pt]{\includegraphics[width=\textwidth]{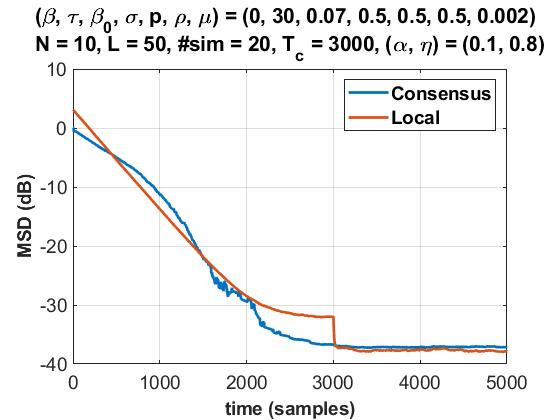}}{\scriptsize{High noise.}}
	\end{subfigure}
	\hfill
	\begin{subfigure}[b]{0.23\textwidth}
		\centering		
		\stackunder[5pt]{\includegraphics[width=\textwidth]{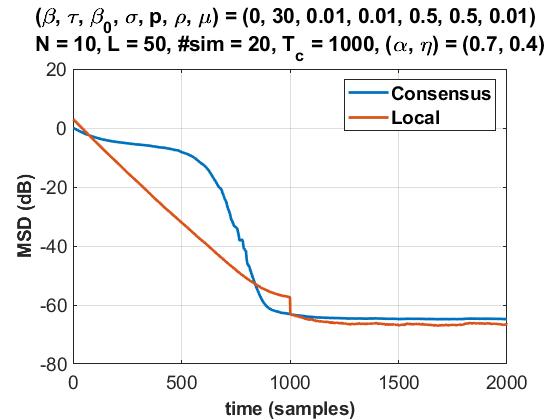}}{\scriptsize{Very low noise.}}
	\end{subfigure}
	\caption[font=small]{
		The DLMS algorithm with DCT-Domain PO.
	}
	\label{fig:POTD}
\end{figure}

In this section, the MSD serves as the performance measure for simulation output, and results are reported on both node and network levels, named as local and consensus estimations, respectively. The local target for node $i$ is $M_i\vec{\omega}^{opt}$, while the network target is $\vec{\omega}^{opt}$ for all nodes. To ensure impartial comparisons, the local MSD is normalized by multiplying it with $\frac{L}{\sum_j D_i(j)}$, where $L$ represents the length of $\vec{\omega}^{opt}$ and $D_i(j)$ is uniformly generated by probability of $\rho$. The combination strategy employed is averaging. During simulation, each link connecting the $N$ nodes is uniformly generated by probability of $p$, and changes in $\eta$ and $\alpha$ are modeled as step functions applied at $t=T_c<T$. In the measurement step, the $\vec{\ab}_{i,t}$ values are independently normally distributed, with the variance of the noise ($\nu_i$) assumed to be $\sigma^2$. Results are reported as averages across the specified number of simulations.
Furthermore, we present two distinctive scenarios for partial observation in our study. The first scenario pertains to partial observation in the time domain ($T=\mathbb{I}$) and is depicted in Fig. \ref{fig:DE}. The second scenario pertains to partial observation in the Discrete Cosine Transform (DCT) domain and is illustrated in Fig. \ref{fig:POTD}. 

Our findings elucidate that by adjusting $\alpha$ and $\eta$, one can manage the flow of information across the network. Additionally, our proposed method can be efficaciously employed in various noise settings. In Fig. \ref{fig:DE}, we investigate two full and partial observation scenarios in the time domain. The results manifestly illustrate that the conventional combination strategy is inadequate. In contrast, our proposed thresholding method (with a fixed threshold level) can enhance information flow across the network even without locally harnessing shared information. Moreover, the adaptive approach can augment each node's estimation by judiciously utilizing the shared information.

Additionally, the effectiveness of our proposed method is demonstrated in fully observable scenarios. Furthermore, in Fig. \ref{fig:POTD}, we corroborate the efficacy of our method in handling DCT-domain partial observations under different noise scenarios by properly tuning the parameters.
\section{Conclusion}
\label{sec:con}
In conclusion, our study presented a framework for analyzing the combination strategy in the Diffusion LMS algorithm where nodes possess partial observations of the target vector, including scenarios with missing information about the vector's support. We proposed a thresholding-based algorithm that can improve information flow across the network and enhance each node's estimation by effectively utilizing shared information. Our simulations demonstrated the effectiveness of the proposed method under various noise scenarios and illustrated its ability to handle partial observability in both time and transform domains. The proposed algorithm successfully mitigated the limitations of the conventional Diffusion LMS algorithm, and the results highlight its potential for applications in practical network settings.
\appendices
\section{Optimal Combination of Partial Observations}
\label{apen}
Given the estimation of node $i$ as $\vec{\omega}_i = M_i\vec{\omega}^{opt}+\vec{e}_i$, the objective is to determine a suitable combination of its neighbors (nodes in $\aleph_i$) in order to minimize the error. This resultant estimation is represented by $\vec{\psi}_i=\sum_{k\in \aleph_i}G_{k,i}.\vec{\omega}_k $, where $G_{k,i}$ signifies the weighting matrix that indicates the effects of $\vec{\omega}_k$ on the $i^{\text{th}}$ node. By defining the error as $J(i)\triangleq\mathop{\mathbb{E}}\{\|\vec{\psi}_i-\vec{\omega}^{opt}\|_2^2\}$, we aim to acquire the optimal weights that will minimize this error.

Throughout the proof procedure, three assumptions are taken into consideration. Firstly, the observation matrix is expressed as $M_i = T'D_iT$, where the unitary transformation matrix, $T$ and its transpose, $T'$, transform to the domain in which the observations' sparsity originates. Secondly, the noise vector, $\vec{e}_i$, adheres to the Orthogonality principle, with zero mean components and variance $\sigma_i^2$. Lastly, the components of $\vec{\omega}^{opt}$ are zero-mean with variance $\sigma_0^2$.
\subsection{Subproblems}
Given the diagonal structure of $M_k=T'\text{diag}(\vec{d}_k)T$, the weighting matrices, $G_{k,i}$, are also assumed to be of the form $G_{k,i} = T'\text{diag}(\vec{g}_{k,i})T$, with the $l^\text{th}$ component ($g_{k,i}(l)$) determining the impact of the corresponding component of $\vec{\omega}_k$ in the transform domain.

With this in mind, the error can be expressed as:
\begin{align}
	J_i=\mathop{\mathbb{E}}\{\|\sum_{k\in \aleph_i}T'\text{diag}(\vec{g}_{k,i})T.( M_k\vec{\omega}^{opt}+\vec{e}_k)-\vec{\omega}^{opt}\|_2^2\}\notag\\
	= \mathop{\mathbb{E}}\{\|[(\sum_{k\in \aleph_i}\text{diag}(\vec{g}_{k,i}) D_k)-\mathbb{I}]	\vec{\Omega}^{opt}\|_2^2\} +
	\mathop{\mathbb{E}}\{\|\text{diag}(\vec{g}_{k,i})\vec{E}_k\|_2^2\},\notag
\end{align}
where, $\vec{\Omega}^{opt}$ and $\vec{E}_k$ represent transformed versions of $\vec{\omega}^{opt}$ and $\vec{e}_k$, respectively.
In consideration of the matter at hand, the error can be segmented into two distinct components: estimation error ($J^{\text{est}}(i)$) and combination error ($J^{\text{comb}}(i)$), which can be expressed as follows:
\begin{align}
	J_i^{\text{est}}&\triangleq \sum_{j=1}^L\sum_{k\in \aleph_i}(g_{k,i}(j))^2\mathop{\mathbb{E}}\{ (E_k(j))^2\}
	=\sum_{j=1}^L\sum_{k\in \aleph_i}(g_{k,i}(j))^2\sigma_k^2,\notag\\
	J_i^{\text{comb}}&\triangleq\sum_{j=1}^L(\sum_{k\in \aleph_i}g_{k,i}(j) d_k(j)-1)^2.\mathop{\mathbb{E}}\{ (\Omega^{opt}(j))^2\}\notag\\
	&=\sum_{j=1}^L(\sum_{k\in \aleph_i}g_{k,i}(j) d_k(j)-1)^2\sigma_0^2.\notag
\end{align}
It is evident that the minimization problem can be broken down into $L$ separate subproblems, each of which can be independently resolved.
\subsection{Optimal Solution to the Subproblems}
Consider a generic node represented as $j$. If the observation index $d_k(j)$ has a value of 0, then there can be no contribution from node $k\in\aleph_i$ during the combination step, leading to no reduction in combination error. Conversely, if $d_k(j)$ has a non-zero value, the error increases by $(g_{k,i}(j))^2\sigma_k^2$. It is therefore reasonable to assign $g_{k,i}(j)=0$ in such cases.  Additionally, in situations where no neighboring nodes exist to estimate the $l^\text{th}$ component of the transformed target signal, its corresponding error will be equivalent to $\sigma_0^2$.

To enhance brevity in the analysis, we can omit unnecessary indices, and assume that $d_k(j)$ is equivalent to $d_k$ and $g_{k,i}(j)$ is equivalent to $c_k$. Additionally, without loss of generality, we can consider $k$ to lie within the range of $1$ to $N$. This allows us to represent the cost function as:
\begin{align}
	J = (\sum_{k=1}^Nc_k d_k-1)^2.\sigma_0^2 + \sum_{k=1}^Nc_k^2\sigma_k^2.\notag
\end{align}
By taking the derivative of the cost function $J$ with respect to $c_l$, and equating it to zero, we obtain:
\begin{align}
	\frac{\partial J}{\partial c_l} = 2d_k(\sum_{k=1}^Nc_k d_k-1).\sigma_0^2 + 2 c_l\sigma_l^2\notag\\
	\xrightarrow{\frac{\partial J}{\partial c_l}=0} c_l = \frac{1-\sum_{k\neq l}c_k d_k}{d_l^2+\lambda_l^{-2}}\times d_l,\notag 
\end{align}
where $\lambda_l^2 = \frac{\sigma_0}{\sigma_l}$ represents the SNR for the given index $l$. It is important to note that if $d_l=0$, it follows that $c_l=0$.
In the special case where $\sigma_k = \sigma$ and $d_k\in \{0,1\}$, the coefficients assume a uniform value of $c_k = c = \frac{1}{m+\lambda^{-2}}$, where $\lambda^2$ is equal to $\frac{\sigma_0^2}{\sigma^2}$ and $m$ represents the count of non-zero $d_k$ values.
\subsection{Solving the System of Coefficients }
\label{apen:sol}
The optimal weights can be obtained by solving for the column vector $\vec{c}$, where $m$ non-zero $d_i$'s are arranged from 1 to $m$, using the following matrix equation:
\begin{equation*}
	\begin{bmatrix}
		d_1^2+\lambda_1^{-2} & d_1.d_2 & \cdots & d_1.d_m \\
		d_2.d_1 & d_2^2+ \lambda_2^{-2} & \cdots & d_2.d_m \\
		\vdots & \vdots & \ddots & \vdots \\
		d_m.d_1 & d_m.d_2& \cdots & d_m^2+\lambda_m^{-2} \\
	\end{bmatrix}\vec{c} =
	\begin{bmatrix}
		d_1 \\
		d_2 \\
		\vdots \\
		d_m \\
	\end{bmatrix}
\end{equation*}
In order to solve the problem, the coefficient matrix is decomposed as the sum of a diagonal matrix ($\Lambda$) and a rank-1 matrix ($B=\vec{\mathbf{1}}.\vec{d}^T$), defined as follows:
\begin{equation*}
	\Lambda \triangleq\text{diag}([\frac{1}{d_1\lambda_1^2},\dots,\frac{1}{d_m\lambda_m^2}]);\;
	B\triangleq \begin{bmatrix}
		1 \\
		1 \\
		\vdots \\
		1 \\
	\end{bmatrix}\begin{bmatrix}
		d_1 & d_2 & \cdots & d_m \\
	\end{bmatrix}
\end{equation*}
This transforms the problem to the form $(\Lambda + B)\vec{c}=\vec{\mathbf{1}}$. The Sherman-Morrison formula can then be used to compute the inverse of $(\Lambda + B)$ as follows:
\begin{equation}
	(\Lambda + B)^{-1} = \Lambda^{-1} -\frac{\Lambda^{-1}\vec{d}.\vec{\mathbf{1}}^T\Lambda^{-1}}{1+\vec{\mathbf{1}}^T\Lambda^{-1}\vec{d}}.\notag
\end{equation}
This yields a solution for the vector $\vec{c}$, with each element computed as:
\begin{equation}
	c_l = d_l\lambda_l^2\times\frac{1-\sum_{k\neq l}d_k(d_k-d_l)\lambda_i^2}{1+\sum_{k=1}^md_k^2\lambda_k^2},\notag
\end{equation}
where $l$ is the index of a typical node\footnote{It can be demonstrated that the utilization of the definition pseudo-inverse can yield equivalent results, even when the observation vector ($\vec{d}$) contains zero values.}.

This outcome highlights the impact of observing the target signal components by each node and its estimation error on the combination step. Furthermore, it is informative to examine instances where the target signal components have equal observability ($d_k=d_0$). In such cases, the following equation holds:
\begin{equation}
	c_l = \frac{d_0\lambda_l^2}{1+d_0^2\sum_{k=1}^m\lambda_k^2},
\end{equation}
where an increase in the relative SNR leads to a higher relative weight among neighboring nodes, as well as a requirement for induced attenuation (or, less commonly, amplification) of the observation.
\subsection{Conventional Combination}
\label{apen:conv}
In this instance, the matrix $G_{k,i}$ is reduced to a scalar weight $g_{k,i}$, resulting in the objective function being rewritten as follows:
\begin{align}
	J(i)&=\mathop{\mathbb{E}}\{\|\sum_{k\in \aleph_i}g_{k,i}(M_i\vec{\omega}^{opt}+\vec{e}_i)-\vec{\omega}^{opt}\|_2^2\}\notag\\
	&=\mathop{\mathbb{E}}\{\|\sum_{k\in \aleph_i}g_{k,i}\vec{E}_i\|_2^2 +\mathop{\mathbb{E}}\{\|(\sum_{k\in \aleph_i}g_{k,i}D_k  -\mathbb{I}
	)\vec{\Omega}^{opt}\|_2^2\}.\notag
\end{align}
Using the above equation, the objective function can be expressed as:
\begin{equation}
	J(i)=\sum_{k\in \aleph_i}g_{k,i}L\sigma_k^2 + \sum_{j=1}^L(\sum_{k\in \aleph_i}g_{k,i}d_k(j)-1)^2\sigma_0^2.\notag
\end{equation}
To simplify the notation, the index $i$ can be eliminated, resulting in a linear system of equations:
\begin{equation}
	(\tilde{\Lambda}+\sum_{j=1}^{L}\vec{d_j}\vec{d_j}^T)\vec{g} = \sum_{j=1}^{L}\vec{d_j},\notag
\end{equation}
where $\tilde{\Lambda}\triangleq L.\text{diag}([\lambda_1^{-2},\dots,\lambda_m^{-2}])$, $\vec{d_j}\triangleq [d_1(j),\dots,d_m(j)]^T$, and $\vec{g}\triangleq[g_{1,i},\dots,g_{m,i}]^T$. According to the Sherman-Morrison formula, the solution $S = (\tilde{\Lambda}+\sum_{j=1}^{L}\vec{d_j}\vec{d_j}^T)^{-1}$ can be recursively found as:
\begin{equation}
	S_r = S_{r-1}+\frac{ S_{r-1}\vec{d_k}.\vec{d_k}^TS_{r-1}}{1+\vec{d_k}^TS_{r-1}\vec{d_k}}; \; S_0 = \tilde{\Lambda}^{-1},S = S_L.
\end{equation}
It is noteworthy to state that the matrix $\tilde{D}=\sum_{j=1}^{L}\vec{d_j}\vec{d_j}^T$ contains elements such as $\tilde{D}_{x,y}$ which demonstrates the cooperation of nodes $x$ and $y$ as neighbors of node $i$. This cooperation is shown through the inner product of their respective observation vectors, denoted as $<\tilde{d}_x,\tilde{d}_y>$. Here, $\tilde{d}_a = [d_a(1), d_a(2),\dots,d_a(L)]^T; a \in\aleph_i$.
Furthermore, each $g_l$ is linked to its aggregative effect, denoted as ($\sum_{j=1}^{L}d_l(j)$), on various components of the combined vector.

As evidenced, unlike the previous problem, finding an optimal solution in this case may not be a simple task, posing as a challenge for estimators that have limited resources. In fact, even in straightforward situations such as a noiseless scenario, an optimal solution may be unattainable.
	\ifCLASSOPTIONcaptionsoff
	\newpage
	\fi

	\bibliographystyle{IEEEbib}
	\bibliography{bibfile}

\end{document}